# Perturbation Method in the Analysis of thin deformed Films and the possible Application

Vjekoslav Sajfert

Technical Faculty "Mihajlo Pupin", Zrenjanin, Yugoslavia

**Abstract**

The perturbation method for the analysis of thin, manifestly deformed films is given. The application of the method the excitons in the film has shown that those have the effective mass essentially depending on the propagation direction. The effects of a mechanical deformation of the film were investigated. It was concluded that the film could serve as an emiter of infrared radiation if the mechanical deformation periodically changes in time.

# Perturbation Method in the Analysis of thin deformed Films and the possible Application

## 1. Introduction

The blank point in the theory of structures with manifestly broken symmetry are the theoretical analyses of thin deformed films. It should be noticed that deformed structures were intensively investigated in two directions. One of them is theory of dislocations, dealing with the problems of partly deformed translationally invariant structures. The substructures where the deformation takes place are very different: one lattice point or several ones, the currves, planes and the parts of total volume. Theory of dislocations was successfully developed for a long time and it gave a series of usefull and interesting results [1-3].

The second domain of research were semiinfinite structures as well as the thin films, deformed only on the boundary surfaces, having the translationally invariant bulk. The main success of these theories is the discovery of surface states and investigation of their properties [4-12].

Thin films deformed in the whole volume are analysed in the papers [13,14]. A set of difference equations.was set up, but during the solution procedure, a continuum transition was performed. As for the other papers treating this subject (papers [l5-19] are some examples), the characteristics of deformed films were studied numerically, but only partially, because only certain experimentally determined values vere commented and explained. It is obvious that in such a manner no general conclusions can be deduced, and every results obtained is only of particular interest.

## 2. The Perturbation Method in the Analysis of thin deformed Films

We shall consider an anisotropic cubic structure with lattice constants $a_x$, $a_y$, and $a_z$. It will be assumed that in this structure the elementary excitations of excitonic type appear. The Hamiltonian of this system, taken in the approximation of the nearest neighbours has the following form

$$H_{id} = \sum_n \Delta B_n^+ B_n + \sum_{s,n} \left( M_{n,n+\lambda_s} + M_{n,n-\lambda_s} \right) +$$
$$+ \sum_{s,n} B_n^+ \left( R_{n,n+\lambda_s} B_{n+\lambda_s} + R_{n,n-\lambda_s} B_{n-\lambda_s} \right) \quad (2.1)$$
$$n \equiv (n_x, n_y, n_z); s = x, y, z; n \pm \lambda_x \equiv (n_x \pm 1, n_y, n_z);$$
$$n \pm \lambda_y \equiv (n_x, n_y \pm 1, n_z); n \pm \lambda_z \equiv (n_x, n_y, n_z \pm)$$

for the ideal, translationally invariant crystal. In the formula (2.1) $\Delta \sim 5\text{eV}$ is the excitation energy of an isolated molecule, $M \sim 0.5\text{eV}$ and $R \sim 0.1\text{eV}$ are the matrix elements of the operator of dipole-dipole interaction and Bose operators $B^+$ and $B$ create and anihilate the excitons. More details on excitonic systems one can find in [20-23].

We shall consider the thin film with boundary surfaces normal to *z*-axis. It will be assumed that film is deformed only along the *z*-direction. In *XY* planes the crystal is infinite and translationally invariant. Index *n* takes the values 0,1,2,3, ..., $N_z$, where $N_z \sim 100-1000$. The layers labeled by $n_z = -1$ and $n_z = N_z + 1$ do not exist.

The main purpose of this poper is to formulate the suitable analytical method for analysis of thin films deformed over the entire thickness. The basic idea of this method is to treat the structural deformation as a perturbation. It is obvious that such approach is applicable only to weakly deformed films. Strongly deformed structures could not be analysed in this way, but, on the other hand, the question is opened whether strongly deformed structures can be considered as a crystals, They are more similar to liquid crystals or to amorphous materials.

In the first part of this work the perturbation method will be exposed. The second part will be devoted to the problem of mechanically deformed optically active crystals. The general method, given in the first part, will be applied to the films deformed by external pressure. The effects which could be expected after removing the pressure will be discussed.

Due to the fact that the film is deformed only along $z$-direction, the quantities $\Delta$, $M$ and $R$ become dependent on the index $n_z$. The following dependence will be assumed:

$$\Delta \to \Delta + \varepsilon(n_z)$$
$$M_{n_x,n_y,n_z;n_x\pm 1,n_y,n_z} = M_x + \alpha_x(n_z)$$
$$M_{n_x,n_y,n_z;n_x,n_y\pm 1,n_z} = M_y + \alpha_y(n_z)$$
$$M_{n_x,n_y,n_z;n_x,n_y,n_z\pm 1} = M_z + \alpha_z(n_z)$$
$$R_{n_x,n_y,n_z;n_x\pm 1,n_y,n_z} = R_x + \beta_x(n_z) \quad (2.2)$$
$$R_{n_x,n_y,n_z;n_x,n_y\pm 1,n_z} = R_y + \beta_y(n_z)$$
$$R_{n_x,n_y,n_z;n_x,n_y,n_z\pm 1} = R_z + \beta_z(n_z)$$
$$\frac{|\varepsilon|}{\Delta} \ll 1; \left|\frac{\alpha_s}{M_s}\right| \ll 1; \left|\frac{\beta_s}{R_s}\right| \ll 1; s=x,y,z$$

Taking into account the symmetry properties of $M$ and $R$ i.e. $F_{n_x,n_y,n_z+1;n_x,n_y,n_z} = F_{n_x,n_y,n_z;n_x,n_y,n_z+1}$, where F stands for M and R, the following is valid:

$$M_{n_x,n_y,n_z+1;n_x,n_y,n_z} = M_{n_x,n_y,n_z;n_x,n_y,n_z+1} = M_z + \alpha_z(n_z+1) \quad (2.3)$$
$$R_{n_x,n_y,n_z+1;n_x,n_y,n_z} = R_{n_x,n_y,n_z;n_x,n_y,n_z+1} = R_z + \beta_z(n_z+1)$$

Finally, due to the absence of layers $n_z = -1$ and $n_z = N_z + 1$ one can write:

$$M_{n_x,n_y,0;n_x,n_y,-1} = M_{n_x,n_y,N_z;n_x,n_y,N_z+1} = 0 \quad (2.4)$$
$$R_{n_x,n_y,0;n_x,n_y,-1} = R_{n_x,n_y,N_z;n_x,n_y,N_z+1} = 0$$

wherefrom it follows:

$$\gamma(0) = \gamma(N_z+1) = 0 \quad (2.5)$$

where $\gamma$ stands for $\alpha$ and $\beta$.

Taking into account (2.2) - (2.4) the Hamiltonian of thin, weakly deformed film can be written as:

$$H_f = H_0 + H_{int} \quad (2.6)$$

where

$$H_0 = H_0' + H_0''$$

$$H'_0 = \sum_{n_x,n_y} \Delta \left( B^+_{n_x,n_y,0} B_{n_x,n_y,0} + B^+_{n_x,n_y,N_z} B_{n_x,n_y,N_z} \right) +$$

$$+ \sum_{n_x,n_y} (2M_x + 2M_y + M_z) \left( B^+_{n_x,n_y,0} B_{n_x,n_y,0} + B^+_{n_x,n_y,N_z} B_{n_x,n_y,N_z} \right) +$$

$$+ \sum_{n_x,n_y} \left( R_x B^+_{n_x,n_y,0} \left( B_{n_x+1,n_y,0} + B_{n_x-1,n_y,0} \right) + R_y B^+_{n_x,n_y,0} \left( B_{n_x,n_y+1,0} + B_{n_x,n_y-1,0} \right) +\right.$$

$$+ R_x B^+_{n_x,n_y,N_z} \left( B_{n_x+1,n_y,N_z} + B_{n_x-1,n_y,N_z} \right) + R_y B^+_{n_x,n_y,N_z} \left( B_{n_x,n_y+1,N_z} + B_{n_x,n_y-1,N_z} \right) +$$

$$\left. R_z B^+_{n_x,n_y,0} B_{n_x+1,n_y,1} + R_z B^+_{n_x,n_y,N_z} B_{n_x+1,n_y,N_z-1} \right)$$

$$H''_0 = \sum_{n_x,n_y} \sum_{n_z=1}^{N_z-1} \Delta B^+_{n_x,n_y,n_z} B_{n_x,n_y,n_z} +$$

$$+ \sum_{n_x,n_y} \sum_{n_z=1}^{N_z-1} (2M_x + 2M_y + M_z) B^+_{n_x,n_y,n_z} B_{n_x,n_y,n_z} +$$

$$+ \sum_{n_x,n_y} \sum_{n_z=1}^{N_z-1} \left( R_x B^+_{n_x,n_y,n_z} \left( B_{n_x+1,n_y,n_z} + B_{n_x-1,n_y,n_z} \right) + \right.$$

$$+ R_y B^+_{n_x,n_y,n_z} \left( B_{n_x,n_y+1,n_z} + B_{n_x,n_y-1,n_z} \right) +$$

$$\left. + R_z B^+_{n_x,n_y,n_z} \left( B_{n_x,n_y,n_z+1} + B_{n_x,n_y,n_z-1} \right) \right) \qquad (2.7)$$

and

$$H_{int} = H'_{int} + H''_{int}$$

$$H'_{int} = \sum_{n_x,n_y} \left( \varepsilon(0) B^+_{n_x,n_y,0} B_{n_x,n_y,0} + \varepsilon(N_z) B^+_{n_x,n_y,N_z} B_{n_x,n_y,N_z} \right) +$$

$$+ \sum_{n_x,n_y} \left( (2\alpha_x(0) + 2\alpha_y(0) + \alpha_z(1)) B^+_{n_x,n_y,0} B_{n_x,n_y,0} + \right.$$

$$+ (2\alpha_x(N_z) + 2\alpha_y(N_z) + \alpha_z(N_z)) B^+_{n_x,n_y,N_z} B_{n_x,n_y,N_z} +$$

$$+ \sum_{n_x,n_y} \left( \beta_x(0) B^+_{n_x,n_y,0} \left( B_{n_x+1,n_y,0} + B_{n_x-1,n_y,0} \right) + \right.$$

$$+ \beta_y(0) B^+_{n_x,n_y,0} \left( B_{n_x,n_y+1,0} + B_{n_x,n_y-1,0} \right) +$$

$$\beta_x(N_z) B^+_{n_x,n_y,N_z} \left( B_{n_x+1,n_y,N_z} + B_{n_x-1,n_y,N_z} \right) +$$

$$\beta_y(N_z) B^+_{n_x,n_y,N_z} \left( B_{n_x,n_y+1,N_z} + B_{n_x,n_y-1,N_z} \right) +$$

$$\left. \beta_z(1) B^+_{n_x,n_y,0} B_{n_x,n_y,1} + \beta_z(N_z) B^+_{n_x,n_y,N_z} B_{n_x,n_y+1,N_z-1} \right)$$

$$H''_{int} = \sum_{n_x,n_y} \sum_{n_z=1}^{N_z-1} \varepsilon(n_z) B^+_{n_x,n_y,n_z} B_{n_x,n_y,n_z} +$$

$$+ \sum_{n_x,n_y} \sum_{n_z=1}^{N_z-1} (2\alpha_x(n_z) + 2\alpha_y(n_z) + \alpha_z(n_z+1) + \alpha_z(n_z)) B^+_{n_x,n_y,n_z} B_{n_x,n_y,n_z} +$$

$$+ \sum_{n_x,n_y} \sum_{n_z=1}^{N_z-1} \left( \beta_x(n_z) B^+_{n_x,n_y,n_z} \left( B_{n_x+1,n_y,n_z} + B_{n_x-1,n_y,n_z} \right) + \right.$$

$$+\beta_y(n_z)B^+_{n_x,n_y,n_z}\left(B_{n_x,n_y+1,n_z}+B_{n_x,n_y-1,n_z}\right)+$$
$$+\beta_z(n_z+1)B^+_{n_x,n_y,n_z}B_{n_x,n_y,n_z+1}+\beta_z(n_z)B^+_{n_x,n_y,n_z}B_{n_x,n_y,n_z-1}\right) \quad (2.8)$$

It is seen from these formulas that the small deformations ε, α and β are included into $H_{int}$ and this part of the Hamiltonian $H_f$ will be treated as a perturbation.

Following the rules of the perturbation theory we shall first solve the eigen value-problem of the Hamiltonian $H_0$. Single-particle excitonic wave function will be taken in the form:

$$|\psi\rangle=\sum_{n_x,n_y,n_z}A_{n_x,n_y,n_z}B^+_{n_x,n_y,n_z}|0\rangle;\ \sum_{n_x,n_y,n_z}\left|A_{n_x,n_y,n_z}\right|^2=1 \quad (2.9)$$

Applying the operator $EB_n-[B_n,H_0]$ to the function (2.9) one obtains the following system of difference equations defining the coefficients A:

$$\left(E-(\Delta+2M_x+2M_y+2M_z)\right)A_{n_x,n_y,n_z}-R_x\left(A_{n_x+1,n_y,n_z}+A_{n_x-1,n_y,n_z}\right)-$$
$$-R_y\left(A_{n_x,n_y+1,n_z}+A_{n_x,n_y-1,n_z}\right)-R_z\left(A_{n_x,n_y,n_z+1}+A_{n_x,n_y,n_z-1}\right)=0$$
$$1\le n_z\le N_z-1$$
$$\left(E-(\Delta+2M_x+2M_y+M_z)\right)A_{n_x,n_y,0}-R_x\left(A_{n_x+1,n_y,0}+A_{n_x-1,n_y,0}\right)-$$
$$-R_y\left(A_{n_x,n_y+1,0}+A_{n_x,n_y-1,0}\right)-R_zA_{n_x,n_y,1}=0$$
$$n_z=0$$
$$\left(E-(\Delta+2M_x+2M_y+M_z)\right)A_{n_x,n_y,N_z}-R_x\left(A_{n_x+1,n_y,N_z}+A_{n_x-1,n_y,N_z}\right)-$$
$$-R_y\left(A_{n_x,n_y+1,N_z}+A_{n_x,n_y-1,N_z}\right)-R_zA_{n_x,n_y,N_z-1}=0$$
$$n_z=N_z \quad (2.10)$$

Keeping in mind that the system is translationally invariant in *XY* planes, we shall look for the solution of the system (2.10) in the following form:

$$A^{(0)}_{n_x,n_y,n_z}=A^{(0)}_{n_z}e^{i(a_xk_xn_x+a_yk_yn_y)} \quad (2.11)$$

After substitution (2.11) into (2.10) we obtain the system of equations:

$$A_1+(\rho-\lambda)A_0=0\,;n_z=0$$
$$A_{n_z+1}+A_{n_z-1}+\rho A_{n_z}=0\,;1\le n_z\le N_z-1 \quad (2.12)$$
$$A_{N_z-1}+(\rho-\lambda)A_{N_z}=0\,;n_z=N_z$$

where

$$\rho=R_z^{-1}\left(\Delta+2(M_x+M_y+M_z)+2(R_x\cos a_xk_x+R_y\cos a_xk_x)-E\right) \quad (2.13)$$

and

$$\lambda=\frac{M_z}{R_z}. \quad (2.14)$$

The secular equation of the system (2.12) is:

$$X_{N_z+1} = \begin{bmatrix} \rho-\lambda & 1 & 0 & 0 & \ldots & 0 & 0 & 0 & 0 \\ 1 & \rho & 1 & 0 & \ldots & 0 & 0 & 0 & 0 \\ 0 & 1 & \rho & 0 & \ldots & 0 & 0 & 0 & 0 \\ & & & \ldots & & & & & \\ 0 & 0 & 0 & 0 & \ldots & 1 & \rho & 1 & 0 \\ 0 & 0 & 0 & 0 & \ldots & 0 & 1 & \rho & 1 \\ 0 & 0 & 0 & 0 & \ldots & 0 & 0 & 1 & \rho-\lambda \end{bmatrix} = 0 \qquad (2.15)$$

The determinant (2.15) can be easily expressed in terms of the second kind Chebyshev's polynomials [24]:

$$T_n = \begin{bmatrix} \rho & 1 & 0 & 0 & \ldots & 0 & 0 & 0 & 0 \\ 1 & \rho & 1 & 0 & \ldots & 0 & 0 & 0 & 0 \\ 0 & 1 & \rho & 0 & \ldots & 0 & 0 & 0 & 0 \\ & & & \ldots & & & & & \\ 0 & 0 & 0 & 0 & \ldots & 1 & \rho & 1 & 0 \\ 0 & 0 & 0 & 0 & \ldots & 0 & 1 & \rho & 1 \\ 0 & 0 & 0 & 0 & \ldots & 0 & 0 & 1 & \rho \end{bmatrix} \equiv \frac{\sin(n+1)\xi}{\sin\xi} \qquad (2.16)$$

$$\rho = 2\cos\xi$$

The connection between $X$ and $T$ is the following:

$$X_{N_z+1} = T_{N_z+1} - 2\lambda T_{N_z} + \lambda^2 T_{N_z-1} = 0. \qquad (2.17)$$

Equating (1.17) to zero one obtains the following transendental equation defining the values of the parameter $\xi$ as well as the values of the parameter $\rho$. In the interval $[0,\pi]$ the equation (2.17) has $N_z+1$ non-zero solution for $\xi$, which will be denoted by $\xi_{\nu_z}$; $\nu_z = 1,2,3,\ldots,N_z+1$.

Since $\rho \to \rho_{\nu_z} = 2\cos\xi_{\nu_z}$ the energies of fexcitons are given by:

$$E_{k_x,k_y,k_z} = \Delta + 2(M_x + M_y + M_z) + 2(R_x \cos a_x k_x + R_y \cos a_x k_x - R_z \cos a_z k_z), \qquad (2.18)$$

where

$$k_z = \frac{\xi_{\nu_z}}{a_z}; \nu_z = 1,2,3,\ldots,N_z+1. \qquad (2.19)$$

The coeficients $A_{n_z} \to A_{n_z}^{k_z}$ can be found in usual way from the system (2.12). The normalized solution is of the form:

$$A_{n_z}^{k_z} = N_{k_z}(-1)^{n_z}(\sin(n_z+1)a_z k_z - \sin n_z a_z k_z) \qquad (2.20)$$

$$N_{k_z} = 2^{1/2}\left(N_x N_y (N_z+1)(1-2\cos a_z k_z + \lambda^2) + 1 - \lambda^2\right)^{-1/2}$$

In accordance with (2.9) and (2.20) the orthonormal wave function of the excitonic system is:

$$\left|\psi_{k_x,k_y,k_z}\right\rangle = \sum_{n_x,n_y,n_z} A_{n_x,n_y,n_z}^{k_x,k_y,k_z} B_{n_x,n_y,n_z}^+ \left|0\right\rangle; \qquad (2.21)$$

$$A_{n_x,n_y,n_z}^{k_x,k_y,k_z} = A_{n_z}^{k_z} e^{i(n_x a_x k_x + n_y a_y k_y)}$$

Since the zero-order energies (2.18) as well as the zero-order wave functions (2.21) are found the further application of perturbation method is well known and a straightforward procedure. We shall quote the first order correction to the excitonic energy (2.18). This correction is:

$$\delta E \equiv \langle \psi_{k_x,k_y,k_z} | H_{int} | \psi_{k_x,k_y,k_z} \rangle =$$
$$= 2((N_z+1)(1-2\lambda\cos a_z k_z + \lambda^2) + 1 - \lambda^2)^{-1} \cdot$$
$$+ \sum_{n_z=1}^{N_z} ( \ (\varepsilon(n_z) + 2\alpha_x(n_z) + 2\alpha_y(n_z) + \alpha_z(n_z) + \alpha_z(n_z+1) +$$
$$+ 2\beta_x(n_z)\cos a_x k_y + 2\beta_y(n_z)\cos a_y k_y)(\sin(n_z+1)a_z k_z - \lambda\sin n_z a_z k_z) -$$
$$- 2\beta_z(n_z)(\sin n_z a_z k_z - \lambda\sin(n_z-1)a_z k_z)) \cdot$$
$$\cdot (\sin(n_z+1)a_z k_z - \lambda\sin n_z a_z k_z)). \quad (2.22)$$

The higher order perturbation corrections can be found in the standard way. The explicit form of these corrections will not be quoted. Instead of that we shall discuss the expression (2.18) defining the zero-order exciton energies in film. This expression differs from the corresponding expression for an ideal structure. The difference consist in the opposite sign of the term proportional to $R_z$.

Assuming $a_x = a_y = a_z \equiv a$ it follows: $M_x = M_y = M_z \equiv M$ $R_x = R_y = R_z = -|R|$ and using the small wave vectors approximation we found from (2.18):

$$E_f = \Delta + 6M - 2|R| + |R|a^2 k^2 (\sin^2\theta - \cos^2\theta), \quad (2.23)$$

where $\theta$ is asimutal angle in $\vec{k}$-space.
The corresponding energy of ideal crystal is given by:

$$E_{id} = \Delta + 6M - 6|R| + |R|a^2 k^2, \quad (2.24)$$

We see that the excitonic gap in ideal structure is less than the corresponding one in the film. The more essential difference lies in the fact that the effective excitonic mass in film depends upon asimutal angle $\theta$, i.e.

$$m_f^*(\theta) = \frac{\hbar^2}{2|R|a^2(\sin^2\theta - \cos^2\theta)}. \quad (2.24)$$

It follows that excitonic effective mass in film depends on the propagation direction. For $\pi/4 < \theta < \pi/2$ effectiv-mass is possitive and consequently excitons have the possitive dispersion. For the propagation directions lying in the intervale $0 \leq \theta < \pi/4$ excitons have negative effective mass and negative dispersion. The excitons cannot propagate on the cone $\theta = \pi/4$ since their effective mass is infinite. The last conclusion can be used for experimental testing of the exposed theory.

**3. Thin Molecular Film under an external Pressure**

In order to illustrate possibilities of the general approach we shall consider the thin molecular film which is deformed by external pressure. It will be assumed that the pressure is applied to the boundary surfaces of the film having the equal magnitude on both surfaces. In this way the lattice constant $a_z$ is symmetrically deformed and becomes the function of index $n_z$, The symmetry plane for $a_z$ is the middle of the film. Consequently, the parabolic form of this function can be used:

$$a_z(n_z) = P + Q\left(n_z - \frac{N_z - 1}{2}\right)^2. \tag{3.1}$$

where $P$ and $Q$ are undetermined coeficients and $N_z$ is an odd number.

The applied pressure maximally changes (decreases) the lattice constant between boundary layers. Consequently, it can be taken:

$$a_z(0) = a_z(N_z - 1) = (1 - \eta)a_z. \tag{3.2}$$

where $a_z$ is the lattice constant of underformed film. z

Assuming that the lattice constant between middle layers remains unchanged by the pressure, i.e.

$$a_z\left(\frac{N_z - 1}{2}\right) = a_z. \tag{3.3}$$

we easily find the function (3.1):

$$a_z(n_z) = a_z\left(1 - \eta\left(1 - \frac{2}{N_z - 1}n_z\right)^2\right). \tag{3.4}$$

It was said earlier that $M$ are matrix element of the operator of dipole-dipole interaction. Consequently it can be taken:

$$M_z \to M_z(n_z) = \frac{M_0}{a_z^3(n_z)} \approx M_z + 3\eta\left(1 - \frac{2}{N_z - 1}n_z\right)^2 M_z + O(\eta^2) \tag{3.5}$$

$$M_z = \frac{M_0}{a_z^3}, \quad \eta \geq 0.$$

Comparing (3.5) to (2.2) we can write:

$$\alpha_z(n_z) = 3M_z\eta\left(1 - \frac{2}{N_z - 1}n_z\right)^2. \tag{3.6}$$

In order to simplify further calculations we shall assume that the deformations of the lattice constants $a_x$ and $a_y$ are negligibly small. This leads to:

$$\alpha_x(n_z) = \alpha_y(n_z). \tag{3.7}$$

The change of matrix elements $R$ will be neglected too, because they are smaller than $M$ for the order of magnitude. Consequently:

$$\beta_x(n_z) = \beta_y(n_z) = \beta_z(n_z) \approx 0. \tag{3.8}$$

It is evident that the pressure cannot change $\Delta$ where from it follows:

$$\varepsilon(n_z) = 0. \tag{3.9}$$

The excitonic energy shift caused by the applied pressure can be calculated from the formula (2.22) after substitutions (3.6) - (3.9). So we obtain:

$$\delta E = \frac{6\eta M_z}{(N_z - 1)^2}\left((N_z - 1)(1 - 2\lambda\cos a_z k_z + \lambda^2) + 1 - \lambda^2\right)^{-1} \cdot$$

$$+ \sum_{n_z=1}^{N_z}\left((N_z - 1 - 2n_z)^2 + (N_z - 3 - 2n_z)^2\right) \cdot \tag{3.10}$$

$$\cdot (\sin(n_z + 1)a_z k_z - \lambda\sin n_z a_z k_z)^2.$$

This formula can be simplified accounting for the fact that $\lambda = M_z / R_z \gg 1$. Neglecting small terms in (2.10) we obtain approximately:

$$\delta E \approx \frac{6\eta M_z}{N_z (N_z - 1)^2} (S_1 - S_2 + S_3) \quad (3.11)$$

where:

$$S_1 = 2(N_z^2 - 4N_z + 5) \sum_{n_z=0}^{N_z} \sin n_z \, a_z k_z$$

$$S_2 = 8(N_z - 2) \sum_{n_z=0}^{N_z} n_z \sin^2 n_z \, a_z k_z \quad (3.12)$$

$$S_3 = 8 \sum_{n_z=0}^{N_z} n_z^2 \sin^2 n_z \, a_z k_z$$

The sums in (3.12) can be easily calculated but exact expressions are very complicated. Taking into account that for real films has to be $N_z \geq 1000$ we can simplify these expressions and their approximate form is:

$$S_1 \approx N_z^3; \quad S_2 \approx 2N_z^3; \quad S_3 \approx \frac{4}{3} N_z^3. \quad (3.13)$$

After substitution of (3.13) into (3.11), where $N_z - 1 \approx N_z$ we finally obtain:

$$\delta E = 2\eta M_z \quad (3.14)$$

For estimation of the magnitude of energy shift (3.14) can be taken $M \approx 0,1 \text{eV}$ and $\eta \approx 0.1$ (the last statement means that for the boundary lattice constants $a_z(0) = a_z(N_z - 1) = \frac{9}{10} a_z$ is valid, i.e. they are shortened by ten percent under the pressure) which lead to

$$\delta E = 0.02 \text{eV} \quad (3.15)$$

To the energy $\delta E$ corresponds the electromagnetic radiation with wavelenght of about 60.000 nm, which is in the domain of far infrared.

The possible application of the last results obtained consists in the following. It is reasonable to expect that the film will illuminate the obtained energy $\delta E$ after removal of the pressure. If the pressure is applied and removed periodically in time the film become a periodical emiter of infrared radiation, i.e. it can serve as a transformator of mechanical energy into electromagnetic, working in a periodical regime.

**4. Conclusion**

The results of the analysis which are carried out in this work can be summarized as follows:
i) The general analytical analysis method of deformed thin films is formulated. This method is based on the idea that the deformation can be treated as a perturhation.
ii) Application of this method to the excitons in thin film has shown that these excitations can change the sign of the effective mass, depending upon the propagation direction.
iii) As a particular case, the influence of external pressure to the excitonic states was analysed. It was shown that the film could work as a periodical transformator of mechanical energy into the electromagnetic if the pressure is applied and removed.

periodically. The electromagnetic waves obtained in this way'should lie in the infrared domain.

**References**

[1] Kadić A. and Edelen D.G.B., A Gange Theory of Dislocations and Disclinations, Berlin-Heidelberg-New York, Springer-Verlag (1983).
[2] Nabarro F.R.N., Theory of Crystal Dislocations, Oxford, Oxford University Press (1967).
[3] Edelen D.G.B., Ann.Phys. **133**, 286 (1981).
[4] Agranovich V.M. and Ginzburg V.L., Crystalooptics Accounting for Space Dispersion and Theory of Excitons, Moscow, Nauka (1979).
[5] Pekar S.I., JETP 38, 1787 (1960)
[6] Sugakov V.I., Fiz. Tverd.Tela, **5**, 2682 (1963)
[7] Bronde V.L. and Romaschik O.K., Uspekhi **9**, 38(1964).
[8] Mills D.L. Phys.Rev. **B3**, 3887(1971)
[9] Roundy V. and Mills D.L., Phys.Rev. **B5**, 1347, (1972)
[10] Binder K. and Hohenberg P.C., Phys.Rev. **B9**, 2194(1974).
[11] Weiner R.A., Phys.Rev. **B8**, 4427(1973).
[12] Sajfert V., Tošić B.S., Marinković M.M. and Kozmidis-Luburić U.F., Physica A **166**, 430 (1990)
[13] Lévy J.C.S. , Surface Sei. Rep. ^, 39 (1981).
[14] Cottam M.G., Tilley D.R. and Žekš B., J. Phys. **C 17**, 1793 (1984).
[15] Bulatov V.L., FTT **26**, 2480 (1984)
[16] Steslicka M., Phys. Lett., **112A**, 234 (1985).
[17] White S.R., Wilkins J.W. and Wilson K.G., Phys. Rev. Lett. **56**, 412 (1986) .
[18] Minkov D., J. Opt. Soc. Am. **A 8**, 306 (1991).
[19] Felske J.D. and Roy P.F., Thin Solid Films **109**, L 113 (1983)
[20] Agranovich V.M., Theory of Excitons, Moscow, Nauka (1978)
[21] Agranovich V.M., JETP **37**, 430(1959).
[22] Davydov A.S., Theory of Molecular Excitons, Moscow, Nauka (1968).
[23] Knox R,, Theory of Excitons, New York, Academic Press (1963)
[24] Gradstein I.S. and Rhyszik I.M., Tables of Integral, Sums, Series and Products, Moscow, FIZMATGIZ (1962) p.1046.